\documentclass[11pt]{JHEP3}
\usepackage{graphicx}
\usepackage{epsfig}
\usepackage{amssymb}
\usepackage{amsmath}
\input epsf.tex

\newcommand{\lsim}{\mbox{\raisebox{-1.ex}{$\stackrel
      {\textstyle<}{\textstyle\sim}$}}}

\newcommand{\del}{\ensuremath{\partial}}

\newcommand{\half}{\ensuremath{\frac{1}{2}}}

\newcommand{\be}{\begin{equation}}
\newcommand{\ee}{\end{equation}}
\newcommand{\bea}{\begin{eqnarray}}
\newcommand{\eea}{\end{eqnarray}}
\newcommand{\ba}{\begin{eqnarray}}
\newcommand{\ea}{\end{eqnarray}}

\newcommand{\ns}{\normalsize}

\title{The fermion spectrum in braneworld collisions}

\author{
  Paul.M.Saffin$^{1}$\footnote{email: paul.saffin@nottingham.ac.uk} and
  Anders Tranberg$^{2,3}$\footnote{email: anders.tranberg@oulu.fi},\\
  $^1${\it\ns School of Physics and Astronomy, University of Nottingham}\\
  {\it\ns University Park, Nottingham NG7 2RD, United Kingdom.}\\  
  $^2${\it\ns DAMTP, University of Cambridge} \\
  {\it \ns  Wilberforce Road, Cambridge, CB3 0WA, United Kingdom.}\\
  $^3${\it\ns Department of Physical Sciences, University of Oulu,} \\
  {\it \ns  90014 Oulu, Finland.}
}


\preprint{DAMTP-2007-??}

\date{}

\maketitle

\abstract{In braneworld collisions fermions originally localised on
one brane can be transferred to another brane, or to a space-time boundary. By
modelling branes as scalar field kinks we bounce them off
boundaries and study resulting effects according to a braneworld observer.
Extending on our previous work, we numerically compute the spectrum of excitations of
fermion modes localised on the brane and boundary, in terms of the momentum $k$
along the brane dimensions. We find that the spectrum depends strongly
on collision velocity and fermion-scalar coupling. Also, high-momentum
modes tend to ``fall off'' the kinks and become delocalised radiation.}

\begin{document}



\section{Introduction\label{sec:introduction}}

In braneworld scenarios our Universe is pictured as a hypersurface living
in a higher dimensional spacetime, with our standard model particles bound
to the hypersurface. In string theory the natural choice for such hypersurfaces
are the D-branes \cite{Johnson:2003gi}, or the boundaries of spacetime \cite{Horava:1996ma}.
However, the braneworld idea was first proposed in the context of field theory models, using
topological defects to construct bound states of scalar fields
on their world-volume \cite{Rubakov:1983bb}. 

In condensed matter systems, it is well known that fermions can be
trapped by impurities in the system, in the sense that the lowest
energy states are localised around these impurities (see, for instance
\cite{Niemi:1984vz}). In a similar way,
topological defects may lead to localised fermion states
\cite{Jackiw:1975fn,Jackiw:1981ee} and one can construct the matter sector of the standard
of the standard model on such defects \cite{Volkas:2007vp}.
\footnote{Note, that there are other
dynamical ways of ensuring localisation to the brane, for instance \cite{Dvali:1996xe}.}.

Multiple branes may exist along the extra dimensions, and so it is
likely that brane-brane collisions will take place. This has prompted
various scenarios for the creation of the cosmic microwave background
anisotropies \cite{Khoury:2001wf}, reheating \cite{Takamizu:2004rq} and baryogenesis.
In the latter
case, the baryon asymmetry is created as the localised
fermions carried by the colliding branes are transferred in a
CP-asymmetric way. 

In \cite{us}, we extended the work of \cite{Maeda}, studying
collisions of branes
represented by topological defects in 1+1 dimensional $\phi^{4}$
theory (a kink), on which fermions were localised. We found that
fermions are indeed transferred in kink-anti-kink collisions, but also
when bouncing a kink off a boundary. These boundaries
can themselves be thought of as branes \cite{Horava:1996ma}.
The final state resulting from kink-boundary and kink-kink collisions
was rather sensitive to the collision velocity and the scalar-fermion coupling,
with an oscillatory rather than monotonic dependence for the number of
transferred fermions \cite{Maeda,us}.

In \cite{us, Maeda}, only the zero momentum fermion modes were considered, and
we now extend the treatment further by including 
the bound-states of fermions with non-zero momenta along the brane dimensions. 
The system we simulate is that of a domain wall in five dimensions colliding with
a (parallel) spatial boundary, we are then able to evaluate the particle spectrum
of fermions resulting from such a collision, as viewed by braneworld observers on
either the brane or boundary.


\section{Scalar-fermion model in 4+1 dimensions\label{sec:model}}

We will model the brane collisions in terms of kinks of a real scalar
field $\phi$ in 4+1 dimensions.
Coupled to this field is a Dirac fermion $\psi$ which, as we shall see, allows
for a bound state localised on the kink; these correspond to the braneworld fields.
The action is 
\bea
\label{eq:action}
S_{\rm bulk}&=&-\int dt\,dz\,d^{3}x\left[ \frac{1}{2}\partial_{\mu}\phi\partial^{\mu}\phi
                    -i\bar{\psi}\gamma^{\mu}\partial_{\mu}\psi
                    +\frac{\lambda}{4}\left(\phi^{2}-1\right)^{2}
                    -ig\phi\bar{\psi}\psi\right],
\eea
to which we add a boundary term
\bea
\label{eq:actionBndry}
S_{\rm boundary}&=&-\int dt\,d^{3}x   \left[ \sqrt{\frac{\lambda}{2}}\left(\frac{1}{3}\phi^3-\phi\right)
                               -\frac{i}{2}\bar\psi\psi\right]_{z=0}.
\eea
For a discussion of this term see \cite{us}, \cite{Antunes:2003kh}.
We will use the conventions
\bea
\eta^{\mu\nu}=diag(-1,1,1,1,1),\quad
\{\gamma_\mu,\gamma_\nu\}=2\eta_{\mu\nu},\quad
\bar\psi=\psi^\dagger\gamma_0,
\eea
and we have singled out the $x^{5}=z$ direction which will be into
                               the bulk, orthogonal to the
                               3-dimensional branes, all of which will
                               be assumed to be parallel to each other.
In 4+1 dimensional spacetime, the fermions are 4-component complex
fields, and we employ the chiral representation, for which
\ba
\gamma_0=\left(
\begin{array}{cc}
0  & -i \\
-i &  0 
\end{array}
\right),~~~
\gamma_J=\left(
\begin{array}{cc}
0 & i\sigma^j \\
-i\sigma^j & 0 
\end{array}
\right),~~~
\gamma_z=-i\gamma_0\gamma_1\gamma_2\gamma_3=\left(
\begin{array}{cc}
-1 & 0 \\
0  & 1 
\end{array}
\right).
\ea
with $\sigma^{j}$ the Pauli matrices. We can then split the spinors as
\ba
\Psi&=&\Psi_L+\Psi_R=
\left(
\begin{array}{c}
\psi_L \\
\psi_R
\end{array}
\right),
\ea
with 
\ba
\Psi_R=\half(1+\gamma_z)\Psi,\qquad \Psi_L=\half(1-\gamma_z)\Psi.
\ea
To match the conventions used in \cite{us}, we write
\ba
\psi_L=\psi_2,\qquad \psi_R=i\psi_1.
\ea
where $\psi_{1,2}$ each have two complex component\footnote{In the
  simpler case of \cite{us}, these were single-component and real.}. 

We assume that the scalar field does not depend on the directions
$\underline{x}={x_{j}}$, $j=1,2,3$ along the brane but only on the
orthogonal direction $z$. In that
case, the scalar equation of motion reduces to
\ba
\ddot{\phi}(t,z)&=&\partial_{z}^{2}\phi(t,z)-\lambda\left(\phi^{2}(t,z)-1\right)\phi(t,z),
\ea
whereas for the fermions we have
\ba
\label{eq:eom1}
\dot\psi_1(t,\underline{x},z)+\sigma^j\del_j\psi_1(t,\underline{x},z)+\del_z\psi_2(t,\underline{x},z)
      -g\phi(t,z)\psi_2(t,\underline{x},z)&=&0,\\
\label{eq:eom2}
\dot\psi_2(t,\underline{x},z)-\sigma^j\del_j\psi_2(t,\underline{x},z)+\del_z\psi_1(t,\underline{x},z)
      +g\phi(t,z)\psi_1(t,\underline{x},z)&=&0.
\ea
We have neglected the fermion back-reaction $ig\langle\bar{\Psi}\Psi\rangle$ on the scalar field
for reasons of simplification.
Our main interest is
in the behaviour of the fermions, and in the setup here back-reaction
would only lead to a small correction to the kink profile, indeed, the zero modes generate no
back-reaction.
Of course,
in the presence of many fermions, such as in a high-temperature
thermal state, such effects cannot be discarded (see
\cite{Aarts:1998td} on how to deal with this).


\subsection{The kink and fermion bound states\label{sec:kink}} 

The static scalar field equation has a kink solution,
\ba
\label{eq:statkink}
\phi_K(z)= \tanh\left(\frac{z-z_{0}}{D}\right),~~D=\sqrt{\frac{2}{\lambda}},
\ea
here centered at $z=z_{0}$ with width $D$. We will operate on the
negative half-axis $z<0$, with the kink incident on a boundary at $z=0$. 

To solve the fermion equation of motion in the background of the static
kink \cite{Randjbar-Daemi:2000cr,Maeda}, we will use a separable ansatz for the fermion field, and write
\ba
\psi_1(t,\underline{x},z)={\cal F}_K(t,z)\chi_1(t,\underline{x}),\qquad
\psi_2(t,\underline{x},z)=0.
\ea
We find that 
\ba
\del_z{\cal F}_K(t,z)+g\phi(t,z){\cal F}_K(t,z)&=&0,\\
\dot{\chi}_1(t,\underline{x})+\sigma^j\del_j\chi_1(t,\underline{x})&=&0.
\ea
The ${\cal F}_K$ is a profile function in the $z$-direction, and is the term responsible for localizing the fermion
on the domain wall,
\ba
\label{eq:fermK}
{\cal F}_K&=&\sqrt{\frac{\Gamma[gD+\half]}{2D\sqrt{\pi}\Gamma[gD]}}\frac{1}{[\cosh(z/D)]^{gD}}=\frac{{\cal N}_K}{[\cosh(z/D)]^{gD}}.
\ea
The equation for $\chi_1$ is that of a right-handed chiral fermion.
Depending on the value of $g$ whole towers of such localised modes
exist but we found that even the second mode contributes
at most $\simeq 10$ percent \cite{us}. We will here restrict our attention to the
lowest mode (\ref{eq:fermK}) only.

We follow the standard quantization procedure and 
expand the component $\chi_1$ in terms of plane waves along the domain wall,
\ba
\chi_1(t,\underline{x})&=&\int\frac{d^3\underline{p}}{(2\pi)^3}\frac{1}{2\omega_p}
\left[ b(\underline{p})e^{ipx}U(\underline{p})
      +d^\dagger(\underline{p})e^{-ipx}V(\underline{p})\right],
\ea
where we require that 
\ba
p^0=\omega_{p}>0\quad |\omega_{p}|=|\underline{p}|.
\ea
The first and second terms in the expansion correspond to positive and negative
energy modes respectively.
Substituting this into the equation of motion for $\chi_1$ we find,
\ba
 b(-\underline{p})[1+\hat{\underline{p}}.\underline{\sigma}]U(-\underline{p})e^{-i\omega t}
-d^\dagger(\underline{p})[1-\hat{\underline{p}}.\underline{\sigma}]V(\underline{p})e^{i\omega
  t}=0,
\ea
with $\hat{\underline{p}}=\underline{p}/\omega_p$. The two terms
must vanish independently, and so
\ba
\label{eq:Udef}
\hat{\underline{p}}.\underline{\sigma}V(\underline{p})=V(\underline{p}),\quad
\hat{\underline{p}}.\underline{\sigma}U(-\underline{p})=-U(\underline{-p}),
\ea
from which we see that
\ba
V(\underline{p})&=&U(\underline{p}).
\ea
So for instance if we have $\underline{p}=(0,0,1)$, then 
\ba
U(p)\propto
\left(
\begin{array}{c}
1\\
0\end{array}
\right),~~~p^{0}>0,~~~\qquad U(p)\propto
\left(
\begin{array}{c}
0\\
1\end{array}
\right),~~~p^{0}<0.
\ea
We have therefore that
\ba
\chi_1(t,\underline{x})&=&\int\frac{d^3\underline{p}}{(2\pi)^3}\frac{1}{2\omega_p}
\left[ b(\underline{p})e^{ipx}U(\underline{p})
      +d^\dagger(\underline{p})e^{-ipx}U(\underline{p})\right],
\ea
where we choose the normalization
\ba
U^\dagger(\underline{p})U(\underline{p})=2\omega_p.
\ea
This last relation fixes the normalization of the operators $b$ and $d$ so that, given
\ba
\Psi(t,\underline{x},z)&=&{\cal F}_K(t,z)
\left(
\begin{array}{c}
0\\
i\chi_1(t,\underline{x})
\end{array}
\right),
\qquad \int\,dz\,{\cal F}_K^\star(t,z) {\cal F}_K(t,z)=1,
\ea
the canonical commutation relations
\ba
\{\Psi_\alpha(t,\underline{x},z),\Psi^\dagger_\beta(t,\underline{x}',z')\}
=\delta_{\alpha\beta}\delta(\underline{x}-\underline{x}')\delta(z-z'),
\ea
reduce to 
\ba
\{\chi_{1\alpha}(t,\underline{x}),\chi^\dagger_{1\beta}(t,\underline{x}')\}
=\delta_{\alpha\beta}\delta(\underline{x}-\underline{x}'),
\ea
and
\ba
\{b(\underline{p}),b^\dagger(\underline{p})\}=\{d(\underline{p}),d^\dagger(\underline{p})\}=(2\pi)^32\omega_p\delta(\underline{p}-\underline{q}).
\ea

We will consider moving kinks, incoming or outgoing with a velocity
$v$, the profile of which are found by applying the appropriate boost
to the static solution (\ref{eq:statkink}),
\ba 
\label{eq:phiKink}
\phi_{v}(t,{\underline{x}},z)&=&\tanh(\gamma(z-z_{0}-vt)/D),
\ea
and correspondingly for the fermion modes
\ba 
\label{eq:boosted1}
\psi_1^{v}(t,\underline{x},z)&=&\sqrt{\frac{\gamma+1}{2}}
{\cal F}_{K}\left(\gamma\left[z-z_{0}-vt\right]\right)\chi_{1}\left(\gamma\left[t-v(z-z_{0})\right],\underline{x}\right),\\
\label{eq:boosted2}
\psi_2^{v}(t,\underline{x},z)&=&\frac{v\gamma}{\gamma+1}\psi_1^{v}(t,\underline{x},z).
\ea


\subsection{The boundary and more fermion bound states\label{sec:boundary}}

The boundary action (\ref{eq:actionBndry}) leads to the boundary conditions
\bea
\label{eq:boundarycondBoth}
\partial_{z}\phi|_{z=0}&=&-\sqrt{\frac{\lambda}{2}}\left(\phi^{2}-1\right)|_{z=0},\\
\psi_1|_{z=0}&=&0.
\eea
These were dubbed $-$BC boundary conditions in
\cite{us} (see also \cite{Antunes:2003kh}). Importantly, the static 
(but not the boosted) kink profile satisfies the boundary condition,
and the kink can hence ``go into'' the wall as it is reflected. Energy
is conserved with the reflected kink having (almost) equal and
opposite velocity to the incoming one, with only a small amount of kinetic energy
lost to scalar radiation.
Fig.
\ref{fig:scalarcollision} shows the kink profile as it hits
the boundary (left) and the evolution of the scalar at the boundary for various
choices of the incident velocity (right). Unsurprisingly, for larger velocity the
kink penetrates further into the wall before bouncing back.
\FIGURE{
\epsfig{file=./pictures/phiprof.eps,width=7cm,clip}
\epsfig{file=./pictures/phivall.eps,width=7cm,clip}
\label{fig:scalarcollision}
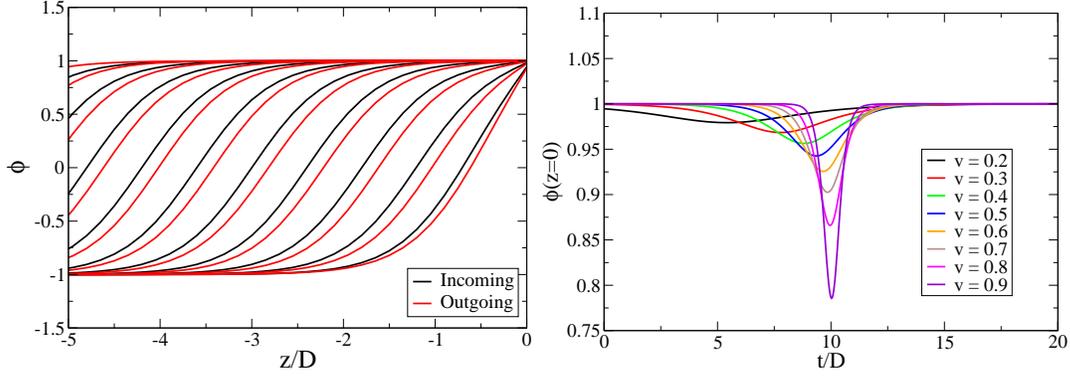
\caption{Left: The incoming kink profiles before, during and after a
  kink-boundary collision, for $v=0.6$. Right: The central value $\phi(t,z=0)$ for
  various collision velocities.}
}

When including the boundary action term (\ref{eq:actionBndry}) it is energetically
favourable for the scalar field to take on the value $1$ (rather than
$-1$) at the boundary \cite{us}, and we take this as the initial 
condition. This is consistent with a kink incoming from the left, which
also has $\phi(z=0)\simeq 1$. 

As well as the kink,
the boundary also carries localised fermion modes and we perform the decomposition
\ba
\psi_1(t,\underline{x},z)=0,\qquad \psi_2(t,\underline{x},z)={\cal F}_B(t,z)\chi_2(t,\underline{x}).
\ea
If the kink is far away, then near the boundary we have that $\phi(z)=1$, and the fermion equation of
motion reduces to
\ba
\del_z{\cal F}_B(t,z)-g{\cal F}_B(t,z)&=&0,\\
\dot\chi_2(t,\underline{x})-\sigma^j\del_j\chi_2(t,\underline{x})&=&0.
\ea
which has the (single) solution
\ba
{\cal F}_B(t,z)=\sqrt{g}\exp(gz)={\cal N}_B\exp(gz),
\ea
for the $z$-dependence and a left-handed chiral fermion for the boundary field.
Note that we are forced to impose the condition $\psi_{1}(z=0)=0$ in order to
find normalisable (and numerically stable) solutions only. 

As a result, the boundary modes have opposite helicity to those on the kink, and 
in order to quantize the system we expand the spinor in terms of these,
\ba
\chi_2(t,\underline{x})&=&\int\frac{d^3\underline{p}}{(2\pi)^2}\frac{1}{2\omega_p}
\left[e(\underline{p})e^{ipx}U(-\underline{p})+f^\dagger(\underline{p})e^{-ipx}U(-\underline{p})\right].
\ea


\subsection{Bogoliubov coefficients\label{sec:bogo}}

For a static kink we have for each mode $\underline{p}$
\ba
\label{eq:kinkModes}
\Psi^{K+}(t,\underline{x},z,\underline{p})=\frac{{\cal N}_K}{[\cosh(z/D)]^{gD}}
\left(
\begin{array}{c}
0 \\
ie^{ipx}U(\underline{p})
\end{array}
\right),\quad
\Psi^{K-}(t,\underline{x},z,\underline{p})=\frac{{\cal N}_K}{[\cosh(z/D)]^{gD}}
\left(
\begin{array}{c}
0 \\
ie^{-ipx}U(\underline{p})
\end{array}
\right),\nonumber\\
\ea
and for the boundary
\ba
\label{eq:boundaryModes}
\Psi^{B+}(t,\underline{x},z,\underline{p})={\cal N}_B\exp(gz)
\left(
\begin{array}{c}
e^{ipx}U(-\underline{p}) \\
0
\end{array}
\right),\quad
\Psi^{B-}(t,\underline{x},z,\underline{p})={\cal N}_B\exp(gz)
\left(
\begin{array}{c}
e^{-ipx}U(-\underline{p})\\
0
\end{array}
\right).\nonumber\\
\ea
In order to quantize the world-volume fields we need to 
introduce a set of creation/annihilation operators. 
We consider the system in the asymptotic past, long
before the collision, and in the far future, long after the collision. In each of these
eras the world volume fields may be expanded in terms of perturbations around the vacuum,
as in standard quantum field theory. As such we write the field operator in the distant
past as
\ba
\Psi(t,\underline{x},z)&=&
\int\frac{d^3\underline{p}}{(2\pi)^3}\frac{1}{2\omega_p}
\left[ b(\underline{p})\Psi^{K+}_{in}(t,\underline{x},z,\underline{p})
      +d^\dagger(\underline{p})\Psi^{K-}_{in}(t,\underline{x},z,\underline{p})\right]\nonumber\\
 &~&+
\int\frac{d^3\underline{p}}{(2\pi)^3}\frac{1}{2\omega_p}
\left[ e(\underline{p})\Psi^{B+}_{in}(t,\underline{x},z,\underline{p})
      +f^\dagger(\underline{p})\Psi^{B-}_{in}(t,\underline{x},z,\underline{p})\right]\nonumber\\
 &~&+\textnormal{bulk continuum modes},
\ea
suitably boosted to velocity $v$. 
In this way we find that: 
$b^\dagger$ creates positively charged, positive helicity particles on the kink;
$d^\dagger$ creates negatively charged, negative helicity anti-particles on the kink;
$e^\dagger$ creates positively charged, positive helicity particles on the boundary;
$f^\dagger$ creates negatively charged, negative helicity anti-particles on the boundary.
We shall not be concerned with the
bulk continuum modes.


We then let the kink collide with the boundary and move away again. The
resulting fermion field operator can then be re-expanded (in the asymptotic future)
in terms of the same modes, boosted in the opposite
direction\footnote{$\Psi_{in}=\Psi(v_{in})$, $\Psi_{out}=\Psi(v_{out})$, where
in practice we determine $v_{out}$ from the evolution of the kink. We
find that $v_{out}=-v_{in}$ to within $10^{-3}$.}
\ba
\Psi(t,\underline{x},z)&=&
\int\frac{d^3\underline{p}}{(2\pi)^3}\frac{1}{2\omega_p}
\left[ \tilde b(\underline{p})\Psi^{K+}_{out}(t,\underline{x},z,\underline{p})
      +\tilde d^\dagger(\underline{p})\Psi^{K-}_{out}(t,\underline{x},z,\underline{p})\right]\\
 &~&+
\int\frac{d^3\underline{p}}{(2\pi)^3}\frac{1}{2\omega_p}
\left[ \tilde e(\underline{p})\Psi^{B+}_{out}(t,\underline{x},z,\underline{p})
      +\tilde f^\dagger(\underline{p})\Psi^{B-}_{out}(t,\underline{x},z,\underline{p})\right]\\
 &~&+\textnormal{bulk continuum modes}.
\ea


By comparing the $e^{ip.x}$ and $U$ dependence of the initial and final states we find that
the two are related through Bogoliubov coefficients $\alpha_{i}$, $\beta_{i}$ as
\ba
\label{eq:bogo1}
\Psi^{K+}_{in}(t,\underline{x},z,\underline{p})&=& \alpha_{K+}(\underline{p})\Psi^{K+}_{out}(t,\underline{x},z,\underline{p})
                                                  + \beta_{K+}(\underline{p})\Psi^{B-}_{out}(t,\underline{x},z,-\underline{p})
                                                  +b.c.m.,\\
\label{eq:bogo2}
\Psi^{K-}_{in}(t,\underline{x},z,\underline{p})&=& \alpha_{K-}(\underline{p})\Psi^{K-}_{out}(t,\underline{x},z,\underline{p})
                                                  + \beta_{K-}(\underline{p})\Psi^{B+}_{out}(t,\underline{x},z,-\underline{p})
                                                  +b.c.m.,\\
\label{eq:bogo3}
\Psi^{B+}_{in}(t,\underline{x},z,\underline{p})&=& \alpha_{B+}(\underline{p})\Psi^{B+}_{out}(t,\underline{x},z,\underline{p})
                                                  + \beta_{B+}(\underline{p})\Psi^{K-}_{out}(t,\underline{x},z,-\underline{p})
                                                  +b.c.m.,\\
\label{eq:bogo4}
\Psi^{B-}_{in}(t,\underline{x},z,\underline{p})&=& \alpha_{B-}(\underline{p})\Psi^{B-}_{out}(t,\underline{x},z,\underline{p})
                                                  + \beta_{B-}(\underline{p})\Psi^{K+}_{out}(t,\underline{x},z,-\underline{p})
                                                  +b.c.m..
\ea
where we have neglected contributions from the bulk continuum modes
and any additional localised excited modes. 
Given that the mode functions are orthonormal we see that
the coefficients are then simply the overlaps of the mode functions, for instance
\bea
\alpha_{K+}(p,t) = 
\int dz\,d^{3}x\left(\Psi^{K+}_{out}(t,\underline{x},z,\underline{p})\right)^{\dagger}\Psi^{K+}_{in}(t,\underline{x},z,\underline{p}).
\eea


We have that
\ba
\tilde b(\underline{p})&=&\alpha_{K+}(\underline{p})b(\underline{p})+\beta_{B-}(-\underline{p})f^\dagger(-\underline{p}),\\
\tilde d(\underline{p})&=&\alpha^\star_{K-}(\underline{p})d(\underline{p})+\beta^\star_{B+}(-\underline{p})e^\dagger(-\underline{p}),\\
\tilde e(\underline{p})&=&\alpha_{B+}(\underline{p})e(\underline{p})+\beta_{K-}(-\underline{p})d^\dagger(-\underline{p}),\\
\tilde f(\underline{p})&=&\alpha^\star_{B-}(\underline{p})f(\underline{p})+\beta^\star_{K+}(-\underline{p})b^\dagger(-\underline{p}).
\ea


The mode functions are all normalised which means that if the bulk modes are negligible
Bogoliubov coefficients relating the bound-state modes form a unitary matrix. 
This implies that 
\ba
|\alpha_{K-}|^2\simeq|\alpha_{B+}|^2,&\qquad& |\alpha_{B-}|^2\simeq|\alpha_{K+}|^2,\\
|\beta_{K-}|^2 \simeq|\beta_{B+}|^2,&\qquad& |\beta_{B-}|^2 \simeq|\beta_{K+}|^2.
\ea

The final particle number in for instance the $K+$ mode can then be
found in terms of the particle numbers in the initial state
\bea
(2\pi)^{3}2\omega_{p}n^{K+}(p)\delta^{3}(p-q)&=&\langle\tilde b(\underline{p})^{\dagger}\tilde b(\underline{q})\rangle\\\nonumber
         &=&\langle \left(\alpha_{K+}(\underline{p})b(\underline{p})+\beta_{B-}(-\underline{p})f^\dagger(-\underline{p})\right)^\dagger
                    \left(\alpha_{K+}(\underline{q})b(\underline{q})+\beta_{B-}(-\underline{q})f^\dagger(-\underline{q})\right) \rangle
\eea
The expectation number of the particle operator then depends on the choice of initial quantum state. If we
suppose that the state before the collision is in the vacuum of the $b$, $d$, $e$, $f$ operators then
we see that the spectrum of particles after the collision is given by
\bea
\label{eq:vacNum1}
n^{K+}(p)&=&|\beta_{B-}(p)|^2,\\
n^{K-}(p)&=&|\beta_{B+}(p)|^2,\\
n^{B+}(p)&=&|\beta_{K-}(p)|^2,\\
\label{eq:vacNum4}
n^{B-}(p)&=&|\beta_{K+}(p)|^2.
\eea
All that remains now is the calculation of the Bogoliubov coefficients, for this we need to solve the Dirac
equation.


\subsection{Individual modes\label{sec:individual}}
As the equations of motion are linear, we can consider each mode
separately. For a given $\underline{p}$ we have
\ba
\psi_1(t,\underline{x},z,\underline{p})=e^{i\underline{p}\underline{x}}\tilde\psi_1(t,z,\underline{p})U(\underline{p}),\quad
\psi_2(t,\underline{x},z,\underline{p})=e^{i\underline{p}\underline{x}}\tilde\psi_2(t,z,\underline{p})U(\underline{p}).
\ea
Then the equations of motion give us
\ba
\label{eq:eomC1}
\dot{\tilde\psi}_1(t,z,\underline{p})+i|\underline{p}|\tilde\psi_1(t,z,\underline{p})+\del_z\tilde\psi_2(t,z,\underline{p})-g\phi(t,z)\tilde\psi_2(t,z,\underline{p})&=&0,\\
\label{eq:eomC2}
\dot{\tilde\psi}_2(t,z,\underline{p})-i|\underline{p}|\tilde\psi_2(t,z,\underline{p})+\del_z\tilde\psi_1(t,z,\underline{p})+g\phi(t,z)\tilde\psi_1(t,z,\underline{p})&=&0,
\ea
where we have used (\ref{eq:Udef})
\ba
\underline{p}.\underline{\sigma} U(\underline{p})&=&
|\underline{p}|\;\underline{\hat p}.\underline{\sigma} U(\underline{p})
=|\underline{p}|U(\underline{p}).
\ea

We solve the equations of motion numerically, starting from the
initial condition of an incoming boosted kink and the appropriate fermion bound state
\ba
\tilde\psi_1&=&\sqrt{\frac{\gamma+1}{2}}\exp[i\omega\gamma v(z-z_{0})]\frac{{\cal N}_K}{[\cosh(\gamma(z-z_{0}))]^{gD}},\\
\tilde\psi_2&=&\frac{v\gamma}{\gamma+1}\tilde\psi_1,
\ea
for the kink mode and 
\ba
\tilde\psi_1&=&0,\\
\tilde\psi_2&=&{\cal N}_B\exp(gz),
\ea
for the boundary mode. Then after the simulation has run its course we
reexpand on the same modes, but boosted with the velocity of the
outgoing kink as described above. This gives us the Bogoliubov coefficients.

This covers the positive energy modes $\Psi^+$ (\ref{eq:kinkModes})(\ref{eq:boundaryModes}). 
The negative energy modes $\Psi^-$ (\ref{eq:kinkModes})(\ref{eq:boundaryModes})
lead to eqs. (\ref{eq:eomC1}), (\ref{eq:eomC2}) with $|p|\rightarrow-|p|$, 
which amounts to complex conjugating the equation of motion, so we should find that
$|\alpha_{K+}|=|\alpha_{K-}|$, $|\beta_{K+}|=|\beta_{K-}|$, $|\alpha_{B+}|=|\alpha_{B-}|$, $|\beta_{B+}|=|\beta_{B-}|$. 
We checked numerically that this is indeed the case, and so in the following we shall only quote
the absolute values $|\alpha_{K}|^{2}$, $|\alpha_{B}|^{2}$, $|\beta_{K}|^{2}$, $|\beta_{B}|^{2}$.


\section{The fermion spectrum\label{sec:spectrum}}

\FIGURE{
\epsfig{file=./pictures/Allex.eps,width=12cm,clip}
\label{fig:Allplot}
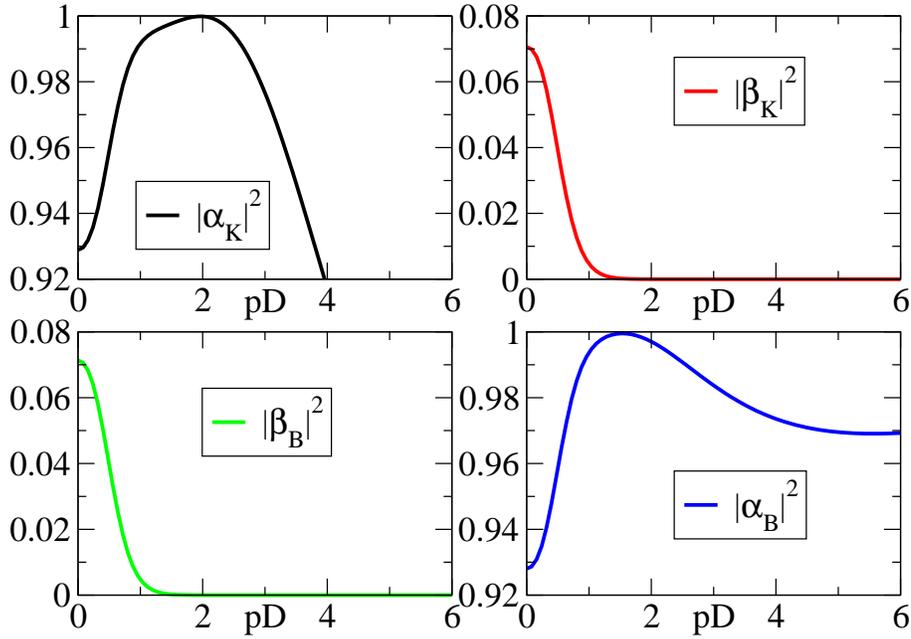
\caption{The $p$-dependence of the Bogoliubov coefficients; of a mode on itself
  $|\alpha_{K,B}|^{2}$, and transfered from one mode to another
  $|\beta_{K,B}|^{2}$. $v=0.6$, $g=2$.} 
}

We show in Fig. \ref{fig:Allplot} the $p$-dependence of the various
Bogoliubov coefficients. In this example, the incident velocity $v$ is $0.6$
and the coupling $g$ is $2$. Throughout the simulations, we use
$\lambda=2$, $D=1$.

If there were no bulk modes, nor any extra excited world-volume modes, we would have that
$|\alpha_K|^2+|\beta_K|^2=1=|\alpha_B|^2+|\beta_B|^2$. Therefore, what Fig. \ref{fig:Allplot}
is telling us is that for $pD\lsim2$ there is very little excitation of the bulk modes, but for
higher momenta the bulk modes become more important.
This is to be expected given that the mass
gap for the bulk fermions is $g$ (where we have the $|\phi|_{bulk}=1$), which for the simulation of Fig. \ref{fig:Allplot} was $g=2$.
Now recall from (\ref{eq:vacNum1})-(\ref{eq:vacNum4}) that the $\beta$ coefficients give us the particle spectrum
for the world-volume fields, if they are in the vacuum state before the collision. We see that the spectrum is
peaked around zero world-volume momentum, with a width in momentum-space given by the inverse of the wall width, $D$.

\subsection{Large $p$\label{sec:largek}}

\FIGURE{
\epsfig{file=./pictures/kmaxg.eps,width=7cm,clip}
\epsfig{file=./pictures/kmaxv.eps,width=7cm,clip}
\label{fig:kmax}
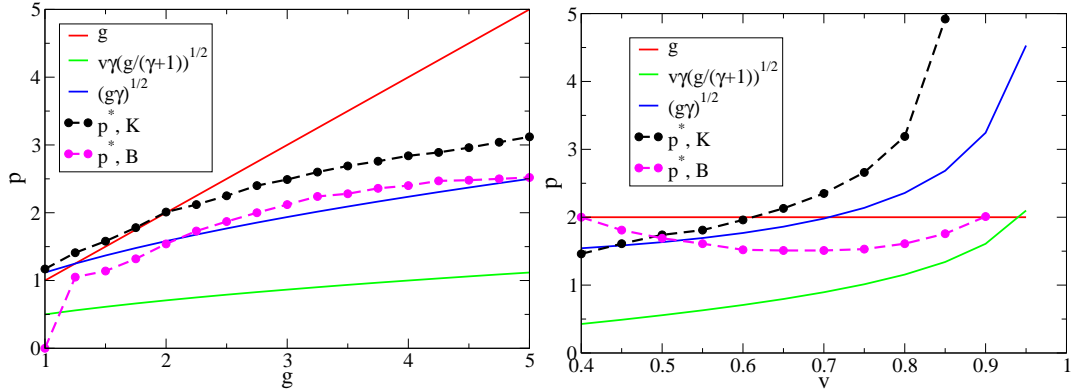
\caption{The momentum $p^{*}$ for which the fermions start to fall off
the kink, compared to simple estimates. $g$-dependence (left) and $v$
dependence (right).}
}
We can understand the behaviour in Fig. \ref{fig:Allplot} by rewriting the equations of motion
(\ref{eq:eomC1}), (\ref{eq:eomC2}) as second order differential
equations, to find
\bea
\label{eq:ddeom}
\ddot{\tilde\psi}_{1}(t,z)=\partial_{z}^{2}\tilde\psi_{1}(t,z)-(p^{2}+g^{2}\phi(t,z))\tilde\psi_{1}(t,z)+g\left(\dot{\phi}\tilde\psi_{2}(t,z)+\partial_{z}\phi\tilde\psi_{1}(t,z)\right),\\\nonumber
\ddot{\tilde\psi}_{2}(t,z)=\partial_{z}^{2}\tilde\psi_{2}(t,z)-(p^{2}+g^{2}\phi(t,z))\tilde\psi_{2}(t,z)-g\left(\dot{\phi}\tilde\psi_{1}(t,z)+\partial_{z}\phi\tilde\psi_{2}(t,z)\right).
\eea
The localisation of the fermion modes is due to the decrease in
$\phi(z)$ and hence mass within the kink. Clearly, when $p^{2}$
dominates $g^{2}\phi^{2}$ as well as the other terms on the right hand side,
the localisation will not be as manifest. 

We can estimate when this will happen by comparing $p^2$ to the size of the other terms. 
In the kink background, we can use (\ref{eq:phiKink})(\ref{eq:boosted1})(\ref{eq:boosted2}) to write
\bea
g^{2}\phi^{2}\tilde{\psi}_{1}<g^{2}\tilde{\psi},\quad g\dot{\phi}\tilde\psi_{1}<\frac{gv^{2}\gamma^{2}}{\gamma+1}\tilde\psi_{2},\quad g\partial_{z}\phi\tilde{\psi}_{1}<g\gamma\tilde{\psi}_{1}.
\eea
Hence, we expect the $p^2$ term in (\ref{eq:ddeom}) to dominate when
\bea
p>g,\quad p>\sqrt{g}\frac{c\gamma}{\sqrt{\gamma+1}},\quad p>\sqrt{g\gamma}.
\eea
Fig. \ref{fig:kmax} (left) shows the $p$ value at which the modes start
to ``fall off'' ($p^{*}$) plotted against $g$, for $v=0.6$. 
This quantity $p^{*}$ is found from the simulations by 
observing where $|\alpha_K|$ starts to decrease.
Keeping in mind to
allow for an overall factor, the functional dependence of $p^{*}$ is
best reproduced by the $\sqrt{g\gamma}$ line, orginating from the spatial
derivative term in (\ref{eq:ddeom}). The $p^*$ derived from the $K$
and $B$ modes are also roughly compatible. In Fig. \ref{fig:kmax} (right) we
fix $g$ and show the dependence on $v$. Again, the behaviour of the
$K$ mode $p^*$ is roughly described by the
$\sqrt{g\gamma}$-dependence. When taking $p^*$ from the $B$ mode,
however, we observe a somewhat different dependence. 
Still, we believe we understand qualitatively the behaviour of the spectrum at
high $p$. Our main focus will be the transfer between $B$ and $K$
modes, whih manifests itself at small $p$.




\subsubsection{Velocity dependence\label{sec:vdep}}

\FIGURE{
\epsfig{file=./pictures/KBvdepg2.eps,width=7cm,clip}
\epsfig{file=./pictures/KBvdepg4.eps,width=7cm,clip}
\label{fig:vdep}
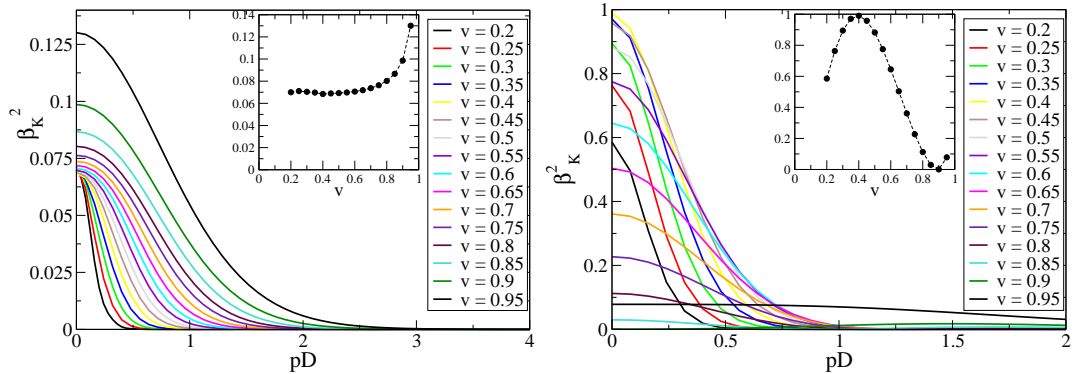
\caption{The velocity dependence of the $|\beta_{K,B}|$-spectrum for $g=2$ (left)
  and $g=4$ (right). Overlaid the $v$-dependence of the $p=0$ mode.}
}
As the incident velocity is increased, the amplitude of the $\phi(z=0)$ dip
  at the collision (Fig. \ref{fig:scalarcollision}, right)
  increases as the kink is able to penetrate deeper into the wall. This
  has a significant impact on the fermion spectrum, as can be seen in
  Fig. \ref{fig:vdep}. For $g=2$, the width of the distribution
  increases as $v$ goes from $0.2$ to $0.5$, beyond which also the
  amplitude starts to grow. We note that this is simply the behaviour
  of the zero-mode \cite{Maeda,us}. For $g=4$ the picture becomes somewhat more
  complicated, in that for $v<0.4$, the distribution grows and widens,
  but then shrinks back to very small values for $v>0.4$. Hence simple
  kinematics does not explain the behaviour. The dependence on
  $v$ is oscillatory rather than monotonic.


\subsubsection{Coupling dependence\label{sec:gdep}}

\FIGURE{
\epsfig{file=./pictures/KBgdepv04.eps,width=7cm,clip}
\epsfig{file=./pictures/KBgdepv09.eps,width=7cm,clip}
\label{fig:gdep}
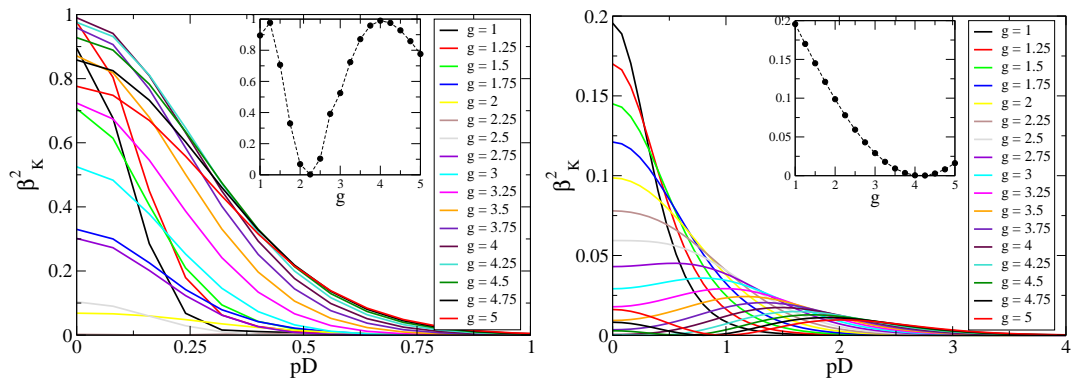
\caption{The coupling dependence of the $|\beta_{K,B}|$-spectrum for $v=0.4$
  (left) and $v=0.9$ (right). Overlaid the $g$-dependence of the $p=0$
  mode. Note the different $p$-ranges in the two plots.}
}

We now fix the velocity and vary the coupling $g$. In \cite{us} we
found that the $k=0$ mode has an oscillatory dependence on $g$, with
amplitude and phase somewhat dependent on $v$. In Fig. \ref{fig:gdep}
(left) we see the spectra for $v=0.4$ at various values of $g$. The
overlay shows the zero-mode oscillation, and to a good approximation,
the spectra are almost Gaussian with the zero mode setting the
amplitude. The width has a separate dependence on $g$. 

In Fig. \ref{fig:gdep} (right) we show the result when fixing $v$ to
$0.9$. Again, we overlay the zero mode result, which sets the
amplitude. But now the spectrum is
(at least for $g>3$) a non-monotonic function of $p$, with maximum away
from $p=0$. Again, the width has a separate dependence on $g$, but
whereas for $v=0.4$ the trend is for the width to shrink with
amplitude, for $v=0.9$ the opposite is the case. Note also the overall
difference in vertical and horizontal scale between the two figures.


\section{Conclusion\label{sec:conclusion}}

Our purpose was to model the particle creation event, as observed by someone
on either brane or boundary, due to the collision between brane and boundary.
We have extended the work of \cite{us} by including fermions with momentum in
the braneworld directions.

We found that a non-trivial fermion spectrum emerges, in general strongly
centered around $p=0$ but at high incident velocities it becomes peaked away from
zero momentum.
The oscillatory dependence, reported in
\cite{Maeda,us}, of the zero momentum coefficients on scalar-fermion coupling
$g$  and velocity $v$ extends to finite $p$. The zero-mode sets the amplitude of the spectrum,
with the width having its own dependence on $v$ and $g$. In the previous study we included
an excited zero momentum bound state as well as the ground state, there we found that
this mode had little impact on the ground state fermion \cite{us}. In principle there will
also be excited modes with non-zero braneworld momentum. While the previous study reveals
that such modes have little impact for low momentum, they may play a more significant role
at higher momentum

Generalising to more complicated
scalar potentials and hence different domain wall profiles is
straightforward, although one may run into trouble when solving
analytically for the bound states. If so these can simply be
found numerically. Such a system has been analyzed for the matter sector of the
standard model \cite{Volkas:2007vp}, although braneworld gauge fields remain problematic. 
With the addition of extra fields we would expect
the spectrum of braneworld particles to thermalize, eventually reaching an equilibrium
distribution.

A natural further extension of this work would be to include multiple
scalar fields (complex, $SU(2)$ doublets) and allow for $C$ and $CP$
symmetry breaking. This of course would be relevant for baryogenesis
models, where branes deposit
asymmetric amounts of matter and anti-matter on the world brane. 

In 3+1 dimensions, similar methods may also be employed specifically for
electroweak baryogenesis, by interpreting the kink as an advancing
bubble wall and calculating reflection and transmission coefficients
for fermions hitting or being hit by the CP-violating wall. We would then have to 
consider a finite temperature state of fermions, and possible include gauge fields 
in the background in addition to the scalar domain wall. 


\subsection*{Acknowledgments}
P.M.S is supported by STFC
and A.T. is supported by STFC Special Programme Grant {\it``Classical
  Lattice Field Theory''}. We gratefully acknowledge the use of the UK
National Cosmology Supercomputer, Cosmos, funded by STFC, HEFCE and Silicon
Graphics.

\bibliographystyle{JHEP}

\end{document}